\documentclass{optica-article}

\journal{opticajournal} 

\articletype{Research Article}

\usepackage{amsmath,amsfonts,bbm,graphicx,color,enumerate,hyperref,xcolor,tikz,ulem,subfigure,physics,threeparttable}
\usepackage{tabularray, textcomp, multirow}
\usepackage{lineno, notes2bib}
\usepackage[Symbol]{upgreek}

\begin{document}

\title{Micromotion compensation of trapped ions by qubit transition and direct scanning of dc voltages}

\author{Woojun Lee,\authormark{1,2,3} Daun Chung,\authormark{1,2} Jiyong Kang,\authormark{1,2} Honggi Jeon,\authormark{2,4} Changhyun Jung,\authormark{2,5,6} Dong-Il ``Dan'' Cho,\authormark{2,5,6} and Taehyun Kim,\authormark{1,2,3,5,7,*}}

\address{\authormark{1}Dept. of Computer Science and Engineering, Seoul National University, Seoul 08826, South Korea\\
\authormark{2}Automation and Systems Research Institute, Seoul National University, Seoul 08826, South Korea\\
\authormark{3}Institute of Computer Technology, Seoul National University, Seoul 08826, South Korea\\
\authormark{4}Dept. of Physics and Astronomy, Seoul National University, Seoul 08826, South Korea\\
\authormark{5}Inter-university Semiconductor Research Center, Seoul National University, Seoul 08826, South Korea\\
\authormark{6}Dept. of Electrical and Computer Engineering, Seoul National University, Seoul 08826, South Korea\\
\authormark{7}Institute of Applied Physics, Seoul National University, Seoul 08826, South Korea}

\email{\authormark{*}taehyun@snu.ac.kr} 


\begin{abstract*} 
Excess micromotion is detrimental to accurate qubit control of trapped ions, thus measuring and minimizing it is crucial. In this paper, we present a simple approach for measuring and suppressing excess micromotion of trapped ions by leveraging the existing laser-driven qubit transition scheme combined with direct scanning of dc voltages. The compensation voltage is deduced by analyzing the Bessel expansion of a scanned qubit transition rate. The method provides a fair level of sensitivity for practical quantum computing applications, while demanding minimal deviation of trap condition. By accomplishing compensation of excess micromotion in the qubit momentum-excitation direction, the scheme offers an additional avenue for excess micromotion compensation, complementing existing compensation schemes.
\end{abstract*}

\section{Introduction}

Trapped ions are considered one of the most promising platforms for realizing a quantum computer \cite{haffner_quantum_2008,bruzewicz_trapped-ion_2019}. They possess multiple advantages including narrow linewidths resulting in long coherence time \cite{wang_single_2021}, well-studied methods of controlling lasers employed from traditional and modern optics \cite{nagerl_laser_1999,hussain_ultrafast_2016,vrijsen_efficient_2019}, and intrinsic reproducibility as natural identical particles \cite{buluta_natural_2011}. Additionally, the ions are trapped by electric fields, which enables tight trapping and crystallization of them \cite{nagerl_coherent_1998}, yet their motional coherence suffers from even a minor presence of undesired electric fields.

In Paul traps, ions are trapped in static (dc) and radio frequency (rf) electric fields. In this configuration, any ion displacement from the rf equilibrium point causes a driven oscillation of the ion at the trapping rf frequency, which is known as an excess micromotion \cite{berkeland_minimization_1998}. The motion poses a detrimental effect within a quantum computing platform that employs trapped ions, due to the system's reliance on motional quantum states for its quantum entangling gates \cite{cirac_quantum_1995,sorensen_entanglement_2000}. Any additional motion would not only decrease trap lifetime, but also reduce the interaction strength with cooling and control lasers and cause heating of the ion, inevitably amplifying quantum decoherence \cite{devoe_observation_1996,poyatos_characterization_1998,turchette_deterministic_1998,rohde_sympathetic_2001,schulz_sideband_2008,poschinger_coherent_2009}. A displacement can occur either during the initial setup before the compensation electrode voltages are optimized to minimize excess micromotion, or at any time afterwards through laser-induced charging of materials in the trap structure, which generates an unwanted electric field at the ion's position \cite{wang_laser-induced_2011, harlander_trapped-ion_2010, harter_long-term_2014, doret_controlling_2012, hong_new_2017}.

Due to its importance, multiple methods to detect and cancel the excess micromotion, or displacement from the rf null have been reported. These methods include measuring the correlation between the phase of the trap rf field and the fluorescence \cite{blumel_chaos_1989,berkeland_minimization_1998,allcock_implementation_2010,shu_heating_2014,pyka_high-precision_2014,keller_precise_2015,zhukas_direct_2021}, measuring a fluorescence spectrum of resolved carrier and rf sideband transitions for a repumping laser or a cooling laser to minimize the sideband \cite{berkeland_minimization_1998,peik_sideband_1999,doret_controlling_2012,vittorini_modular_2013,shu_heating_2014,keller_precise_2015,goham_resolved-sideband_2022}, using parametric excitation while modulating the trap rf voltage with the secular frequency \cite{ibaraki_detection_2011,mount_single_2013,keller_precise_2015,auchter_industrially_2022}, and measuring the ion position while altering the amplitude of the confining dc or rf potential to minimize the displacement \cite{berkeland_minimization_1998,brown_loading_2007,doret_controlling_2012,saito_measurement_2021,auchter_industrially_2022}.

In this paper, we demonstrate a method to detect the excess micromotion by a single qubit rotation while directly scanning the dc voltage, leveraging the existing scheme for qubit control experiments. Even though it employs the conventional principle of exploiting the change in transition rates caused by excess micromotion, the method can be implemented by using only the fixed frequency of either the carrier transition or an rf sideband transition, diminishing the need for scanning in the frequency or sideband dimension in addition to iteratively adjusting the compensation voltage as in the sideband minimization method. Additionally, our method minimizes deviation of the trap environment, such as alternating between two different amplitudes of the trap rf voltage, adjusting the frequency or intensity of the repumping or cooling laser beam, or parametrically exciting the motion of the ion, all of which could potentially compromise the stability of the trapping during the measurement. Furthermore, the method does not employ the cooling laser as in the photon correlation method, avoiding the spectral interference with the scattering of the laser. These features allow for the direct application of the method to frequently encountered situations in ion trap system where the excess micromotion along the motion-controlled axis for quantum gates cannot be easily detected by the conventional methods.

\section{Modification of transition probability by excess micromotion}
\label{sec_bessel}

The amount of excess micromotion can be detected by measuring the qubit transition probability while scanning the dc voltage of the trap to tune the equilibrium position of the ion. Consider an ion trapped by an rf pseudopotential $\phi_\mathrm{rf,x} = m\omega_x^2 x^2 / 2e$ along the $x$-axis, where a voltage change of the controlled dc electrode $\mathrm{\delta} V_\mathrm{dc}$ creates an electric field at the ion position by $\mathrm{\delta} E_{\mathrm{dc},x}$, where  $\omega_x$ is the secular frequency in $x$-axis, $m$ is the mass of the ion, and $e$ is the unit charge. Using the ratio of the differential electric field to the differential voltage of the controlled dc electrodes, $K_x=\mathrm{d}E_{\mathrm{dc},x}/\mathrm{d}V_\mathrm{dc}$, the displacement by this field in the pseudopotential can be written as $u_{\mathrm{dc},x} = e K_x \mathrm{\delta} V_\mathrm{dc} / m\omega_x^2$, and the total displacement including the displacement by a stray field $E_{\mathrm{stray},x}$ is expressed as
\begin{equation}
\label{eq_u0}
    u_{0x} = u_{\mathrm{dc},x}+u_{\mathrm{stray},x} = e ( K_x \mathrm{\delta} V_\mathrm{dc} + E_{\mathrm{stray},x}) / m\omega_x^2.
\end{equation}

\begin{figure}[ht]
 \centering
 \includegraphics[width=1\textwidth]{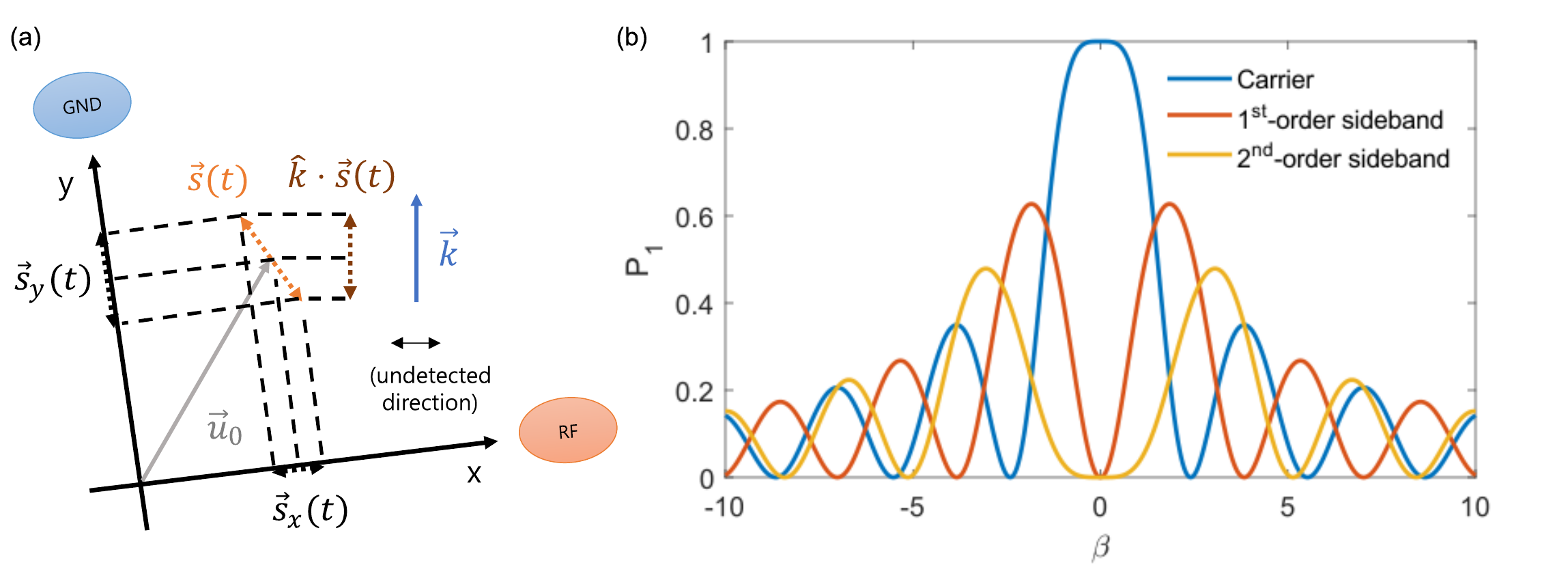}
 \caption{(a) Schematic representation of two-dimensional excess micromotion and probe beam momentum. For simplicity of explanation, it is assumed that the principal axes are not tilted with respect to the x and y axes. The ground and rf electrodes are drawn as examples. (b) Modification of transition probability by excess micromotion for motional sidebands.}
 \label{fig_bessel}
\end{figure}

To estimate the magnitude of the field to be compensated at the ion position, the transition probability for a qubit flipping is measured while the voltages of the controlled dc electrodes are varied. Excess micromotion occurs along the direction of the oscillating electric field generated by the rf voltage at the displaced position, which is determined by the spatial arrangement of ground and rf electrodes as shown in Fig. \ref{fig_bessel}(a), where the ion is displaced by $\textbf{u}_0 = (u_{0x}, u_{0y})$ from the rf null. The instantaneous ion displacement $\textbf{u}(t) = (u_x(t), u_y(t))$, including the secular motion with amplitudes $u_{1x}$, $u_{1y}$ at frequency $\omega_x, \omega_y$ and the micromotion at frequency $\omega_\mathrm{rf}$, can be expressed as,
\begin{equation}
    u_i(t) \approx [ u_{0i} + u_{1i} \cos(\omega_i t + \phi_i)] \biggr[ 1 + \frac{q_i}{2}\cos(\omega_\mathrm{rf} t) \biggr],
\end{equation}
where $i$ can be either $x$ or $y$, and $q_{x,y}$ are trap parameters known as the $q$-parameters along the principal axes, which depend on the trap geometry and are proportional to the magnitude of the pseudopotential \cite{leibfried_quantum_2003,berkeland_minimization_1998}. Note that $q_x \cdot q_y < 0$ because the ground and rf electrodes have opposite signs of voltage with respect to the voltage at the ion's equilibrium position. If the ion is sufficiently cooled down, the secular motion amplitudes $u_{1x}$ and $u_{1y}$ become negligible, and the pure excess micromotion term can be written as
\begin{equation}
\label{eq_mm}
    \textbf{s}(t) = \textbf{u}(t) - \textbf{u}_0 \approx \biggr( \frac{u_{0x} q_x}{2}, \frac{u_{0y} q_y}{2} \biggr) \cos(\omega_\mathrm{rf} t) = \textbf{s}_0 \cos(\omega_\mathrm{rf} t).
\end{equation}

For detection of this excess micromotion, we utilize an atom-light interaction of the ion with a transition laser of the momentum $\textbf{k}$ (or the momentum difference of $\textbf{k}=\textbf{k}_2-\textbf{k}_1$ in the case of Raman transition with momenta $\textbf{k}_1$ and $\textbf{k}_2$), where the excess micromotion component projected in the direction of $\textbf{k}$ can be detected. When the transition laser field is represented by $E(t)$ in the lab frame, the oscillating ion experiences the phase modulation of the field by $\textbf{k} \cdot \textbf{s}(t) = \beta \cos(\omega_\mathrm{rf} t)$  and therefore the electronic transition of the ion is driven by the modified electric field
\begin{equation}
\label{eq_field}
E'(t) = E(t) \exp \left[ \mathrm{i} \beta \cos(\omega_\mathrm{rf} t) \right]
\end{equation}
with the modulation depth of
\begin{equation}
\label{eq_beta}
    \beta = \textbf{k} \cdot \textbf{s}_0.
\end{equation}
Note that the modulation depth defined here can be not only non-negative but also negative, contrary to the usual definition. The modulation can be expanded with the Bessel function of the first kind $J_n (x)$ as,
\begin{equation}
\label{eq_phase}
\exp \left[ \mathrm{i} \beta \cos(\omega_\mathrm{rf} t) \right] = \sum_{n=-\infty}^{\infty} J_n (\beta) \exp \left[ \mathrm{i} n \left( \omega_\mathrm{rf} t + \uppi/2 \right) \right].
\end{equation}
The modulated field for transition directly affects the Rabi frequency of the trapped ion with the following interaction Hamiltonian,
\begin{equation}
    \hat{H}_\mathrm{I} = \frac{\hbar}{2}
    \begin{pmatrix}
    0 & \Omega_\mathrm{o}\\
    \Omega_\mathrm{o} & 0
    \end{pmatrix},
\end{equation}
replacing the Rabi frequency $\Omega_\mathrm{o} = \mu |E(t)|/\hbar$ with the modified Rabi frequency $\Omega_\mathrm{o}'$, where $\mu$ is the transition dipole moment.
The resulting time evolution of the ion's quantum state then becomes
\begin{equation}
\ket{\psi (t)} = \cos (\Omega_\mathrm{o}' t/2) \ket{0} + \sin (\Omega_\mathrm{o}' t/2) \ket{1}.
\end{equation}
If the ion interacts with a transition field at a detuning of $n$-th order sideband, $n \omega_\mathrm{rf}$, the time-dependent factor $\exp (-\mathrm{i}n\omega_\mathrm{rf}t)$, keeps the $n$-th terms only in the expansion of Eq. (\ref{eq_phase}) and the other terms would be time-averaged to vanish. In this case, Eq. (\ref{eq_field}) reduces to $E'(t) = J_n (\beta) E(t) \exp \left[ \mathrm{i} n \left( \omega_\mathrm{rf} t + \uppi/2 \right) \right]$, and the modified Rabi frequency becomes $\Omega_\mathrm{o}' = \mu |E'(t)|/\hbar = J_n (\beta) \Omega_\mathrm{o}$.

If we choose to measure the carrier transition only, for instance, we can extract $n=0$ term from the expansion so the modified Rabi frequency is $\Omega_\mathrm{o}' (\beta) = J_0 (\beta) \Omega_\mathrm{o}$.
The transition probability to $\ket{1}$ state with a $\uppi$-pulse with a pulse time of $t_\uppi = \uppi / \Omega_o$ will then follow
\begin{equation}
\label{bessel}
P_{\ket{1}, n=0}(\beta; t_{\uppi}) = |\bra{1}\ket{\psi(t_\uppi)}|^2 =  \left| \sin \left( \Omega_\mathrm{o}' t_{\uppi} /2 \right) \right|^2 = \left| \sin \left( \uppi J_0 (\beta) /2 \right) \right|^2.
\end{equation}
Here, $\beta = A \mathrm{\delta}V_\mathrm{dc}+B$ can be varied with the controlled dc voltages by $\mathrm{\delta}V_\mathrm{dc}$, where $A=(e/2m)(k_x q_x K_x/\omega_x^2+k_y q_y K_y/\omega_y^2)$ and $B=(e/2m)(k_x q_x E_{\mathrm{stray},x}/\omega_x^2+k_y q_y E_\mathrm{stray,y}/\omega_y^2)$, from Eqs. (\ref{eq_u0}), (\ref{eq_mm}), and (\ref{eq_beta}). Given that $\beta$ can be either positive or negative, $P_{\ket{1}, n}(\beta; t_{\uppi})$ is an even function of $\beta$. Similarly, the transition probability for the $n$-th sideband can be expressed as,
\begin{equation}
\label{eq_rabi}
    P_{\ket{1}, n}(\beta; t_{\uppi}) = \left| \sin \left( \uppi J_n (\beta) /2 \right) \right|^2.
\end{equation}
Calculated transition probabilities for $n=0, 1,$ and $2$ against $\beta$ are plotted in Fig. \ref{fig_bessel}(b).

Specifically for the carrier transition, $P_{\ket{1}, n=0}(\beta; t_{\uppi})$ has a global maximum at $\beta=0$. Scanning the controlled dc voltages while applying a $\uppi$-pulse of the transition each time, the transition probability will have a global maximum at a set of dc voltages whose dc null matches the rf null, which could be directly used to find the compensation voltage for minimization of excess micromotion.

\section{Experimental setup}

\begin{figure}[ht]
 \centering
 \includegraphics[width=1\textwidth]{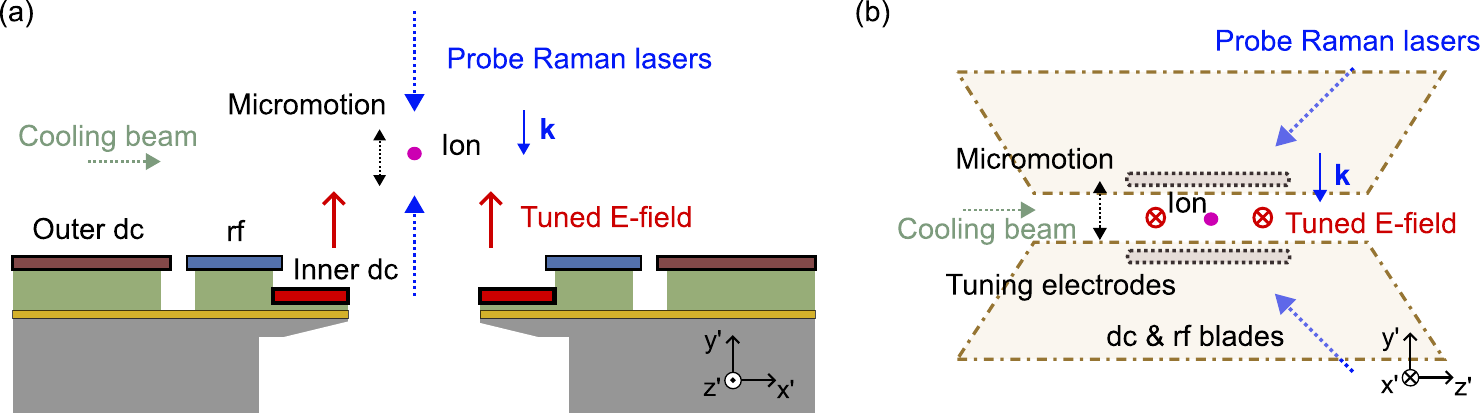}
 \caption{Experimental setup and configuration for (a) the surface and (b) the blade traps, with the directions of the tuned electric fields, excess micromotions, probing Raman lasers and corresponding momenta (momentum differences for Raman lasers) illustrated.}
 \label{fig_setup}
\end{figure}

The suggested method was used as a part of the following two experimental setups for trap calibration: a microfabricated surface trap and a blade trap.

The surface trap (see Fig. \ref{fig_setup}(a)), or the chip trap, is a silicon-based microfabricated chip with dc and rf electrodes, made of aluminum, on the top surface of the chip and coated by gold on top (more detailed descriptions of the chip structure can be found in \cite{jung_microfabricated_2021}). $^{171}$Yb$^+$ ions are trapped on the chip at a height of 80 µm in an ultra-high vacuum of $< 1\times10^{-10}$ mbar ($< 1\times10^{-8}$ Pa). There is a loading slot with a width of 80 µm along the trap axis direction ($z'$-axis in Fig. \ref{fig_setup}(a)), which penetrates the chip. An ion is tightly confined radially in a pseudo-potential generated by rf voltages with a frequency of 22.2 MHz and an amplitude of approximately 200 V, and loosely confined along the trap axial direction by a static potential generated by a set of dc voltages. The voltages are applied to the inner and outer dc electrodes to construct the trapping potential supplied by a digital-to-analog converter (DAC; ADLINK Technology PCI-6208), and the pair of inner dc electrodes are particularly used for tuning of the compensation field. The secular frequencies of the trap along the three principal axes are 1.6 MHz and 1.4 MHz for the two radial directions ($0.64\hat{x}'+0.77\hat{y}'$, $0.77\hat{x}'-0.64\hat{y}'$), and 450 kHz for the axial direction ($\hat{z}'$), respectively. A 369-nm Doppler cooling beam is injected in $\hat{x}'+\hat{z}'$ direction to have non-zero overlaps with all three principal axes. Imaging and state detection of the trapped ion are achieved by collecting the 369-nm fluorescence with a high-NA imaging lens (Photon Gear 15470-S, 0.6 NA) and counting the photons with an electron-multiplied charge-coupled device (EMCCD) or a photomultiplier tube (PMT). For Raman transition from $\ket{0}=\ket{^2 S_{1/2}, F=0, m_F = 0}$ state to $\ket{1}=\ket{^2 S_{1/2}, F=1, m_F = 0}$ state to achieve Rabi oscillation, a 355-nm picosecond pulse laser with a repetition rate of around 120 MHz is split into two beams and separately modulated with acousto-optic modulators (AOM) to become a pair of frequency-locked Raman beams \cite{campbell_ultrafast_2010}. The beams with waists of 2 µm and 15 µm were injected onto the ion in a counter-propagating configuration from the front (above) and back (below) of the chip respectively, as shown in Fig. \ref{fig_setup}(a).

The blade trap, which follows a typical structure, comprises four blade electrodes with a spacing of 460 µm, operated in an ultra-high vacuum (see \cite{jeon_experimental_2023} for more details of the trap). The trap rf voltage has a frequency of 15.3 MHz and an amplitude of 700 V, and the secular frequencies of the trap are 1.25 MHz, 1.28 MHz for the radial directions ($0.41\hat{x}'+0.91\hat{y}'$, $0.91\hat{x}'-0.41\hat{y}'$), and 120 kHz for the axial direction ($\hat{z}'$), respectively. A 369-nm Doppler cooling beam is injected in $\hat{x}'+\hat{z}'$ direction to have non-zero overlaps with the three principal axes. The 369-nm fluorescence is collected by a homemade lens assembly with 0.34 NA and counted by an EMCCD or a PMT for imaging and state detection. Raman transition scheme is equivalent to that of the surface trap, and the Raman beams with a waist of 10 µm are injected in perpendicular configuration as shown in Fig. \ref{fig_setup}(b). Tuning of the dc voltages for stray field compensation is achieved through additionally installed electrodes, supplied by a high voltage power supply (SRS PS350). Note that, in the blade trap, the excess micromotion compensated by the tuning electric field is in a perpendicular direction to the ion displacement due to the rf field profile of the trap.

\section{Result}
\label{sec_result}

\begin{figure}[ht]
 \centering
 \includegraphics[width=0.7\textwidth]{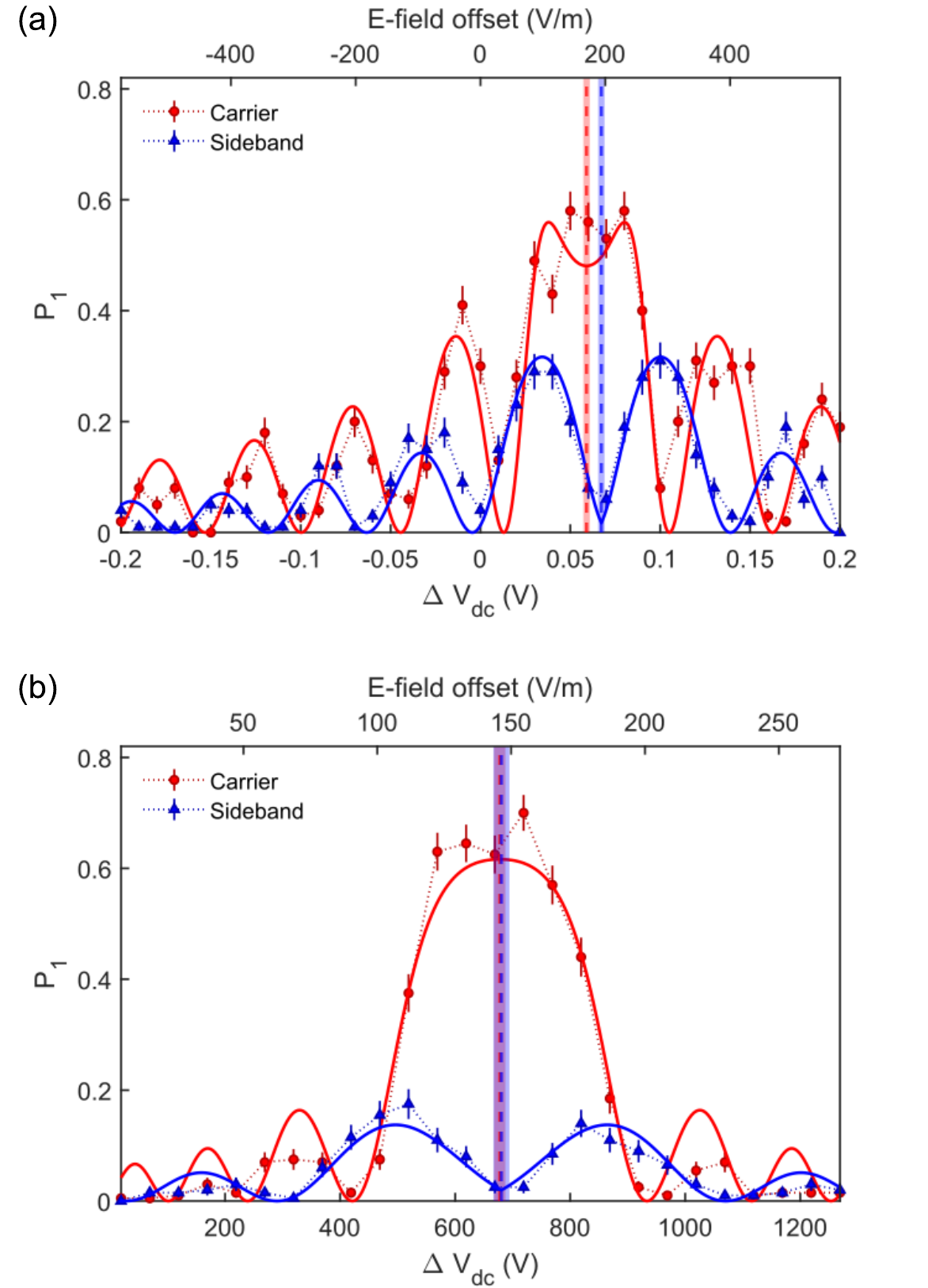}
 \caption{Measurement for (a) the chip trap and (b) the blade trap. The vertical dashed lines present the estimated compensation voltages for the fitting curves. The error bars of the points indicate the standard errors of the measurements, and the 95 \% confidence intervals of the estimated compensation voltages are marked by the shaded regions.}
 \label{fig_scan_raman}
\end{figure}

In both the surface trap and the blade trap, state detection of $\ket{1}$ was conducted after applying near-$\uppi$ pulses to qubits in the initial condition $\ket{0}$, while scanning the dc voltage. The experimental data was fitted to a Bessel-like fitting curve as shown in Fig. \ref{fig_scan_raman}, along with consideration of the practical experimental conditions such as the thermal decoherence, anharmonicity in the trap potentials, intrinsic micromotion, and the actual pulse time which can be shorter or longer than the $\uppi$-pulse time. In order to demonstrate the practical measurement capability, the data was collected only after the typical Doppler cooling which will leave the ion with thermal distribution of phonons. Accordingly, the data was fitted to a modified Rabi oscillation profile accounting for thermal motion, $f_{N_\mathrm{ph}} (a J_n (\beta); t)$, where $N_\mathrm{ph}$ is the average number of thermal phonons and $a$ is a fitting parameter which is proportional to the transition strength, with the consideration of different Rabi frequencies with different phonon numbers \cite{semenin_determination_2022}. The transition probability in Eq. (\ref{eq_rabi}) is then replaced by
\begin{equation}
\label{eq_thermal}
P_1 = f_{N_\mathrm{ph}} \bigr[ a J_n ( b_0 + b_1 v + b_2 v^2  ); t_{p} \bigr]
\end{equation}
with $v = \Delta V_\mathrm{dc}-c$, for fitting parameters $N_\mathrm{ph}, b_0, b_1, b_2$, and $c$, where $t_{p}$ is the pulse time, and $c$ is considered the compensation voltage which is depicted as dashed vertical lines in the plots. The data was obtained for both the carrier transition and the first-order sideband transition of the trap rf frequency, either of which can be used to find the compensation voltage. Fig. \ref{fig_scan_raman}(a) is for the surface trap and (b) for the blade trap.

The estimated sensitivity of the measurement to measured compensation voltage, the corresponding magnitude of the field, and $\beta$, for the two types of trap and transitions could be found in Table \ref{tb_sensitivity}, where sensitivity here is defined as the one-half value of 95\% confidence interval of a curve fitting. The magnitude of the electric field in the surface trap was estimated using a COMSOL Multiphysics® simulation, which was calculated to be 2880 V/m at the ion position per 1 V of inner dc pair, and 0.224 V/m per 1 V of tuning electrodes in the blade trap. The result shows that a sensitivity comparable to the reported levels of detection or compensation sensitivities obtained in practical applications \cite{doret_controlling_2012,vittorini_modular_2013,auchter_industrially_2022} could be achieved by the suggested approach. The compensation voltage can be determined using any of the carrier or sideband transitions, and it requires only a minimum of 10 to 20 data points. Note that the above measurement was carried out as characterization of excess micromotion sufficient for our daily experiments while the uncertainty can be further reduced by more precise and repeated measurements.

\begin{table}
\caption{Estimated sensitivity of the measurement to compensation voltage, stray field, and modulation depth.}
\label{tb_sensitivity}
\centering
\begin{tblr}{
  cells = {c},
  cell{1}{2} = {c=2}{},
  cell{1}{4} = {c=2}{},
  hlines,
}
                     & Surface trap &          & Blade trap &          \\
                     & Carrier      & Sideband & Carrier    & Sideband \\
Voltage (V)          & 0.0018       & 0.0017   & 9.7        & 13.9     \\
Electric field (V/m) & 5.1          & 5.0      & 2.2        & 3.1      \\
$\beta$              & 0.091        & 0.084    & 0.048      & 0.093     
\end{tblr}
\end{table}

The result shows that either of the measurements with the carrier transition and the sideband transition provide decent levels of sensitivity, which can be further improved in uncertainty by combining those two results. Especially, if there exists any ambiguity about the global maximum (minimum) due to multiple local maximums (minimums), comparison of two independent scans with the carrier transition and the sideband transition can allow us to identify the true null position because of the complementary behavior of the carrier and sideband transitions.

\begin{figure}[ht]
 \centering
 \includegraphics[width=1\textwidth]{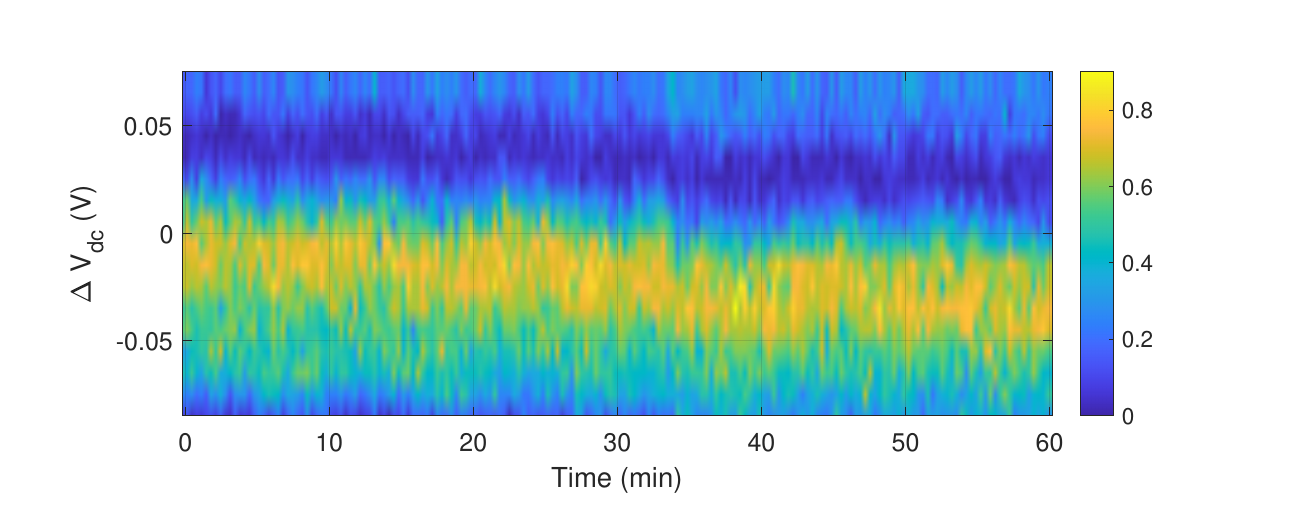}
 \caption{Measurement of stray field drift caused by long-term dielectric charging of the trap chip materials. A set of transition probabilities with varying dc voltages was repeatedly obtained and depicted as each vertical profile.}
 \label{fig_drift}
\end{figure}

Note that, in order to demonstrate that the oscillatory pattern of the measured probability originates from modulation by the Bessel function, we intentionally scanned over a wide range of dc voltage for both carrier transition and the sideband transition, and Fig. 3 (a) shows that the resulting two fitted curves agree well with measurements. Scanning across the extended voltage range inevitably required more time compared to the typical scan period of less than 30 seconds. Consequently, notable drift occurred between the carrier scan and the sideband scan due to the unavoidable delay between two scans and the measured drift is larger than the sensitivity of each scan as can be seen in Fig. \ref{fig_scan_raman}(a). This type of drift due to long-term charging of dielectric materials within the microfabricated trap is relatively well-known \cite{wang_laser-induced_2011, harlander_trapped-ion_2010, harter_long-term_2014, doret_controlling_2012, hong_new_2017} and the amount of drift can also increase over time as shown in Fig. \ref{fig_drift}. On the other hand, the drift rarely occurs in a blade trap as shown in Fig. \ref{fig_scan_raman}(b) because of the macroscopic size of the trap structure.

Before any compensation, the exact $\uppi$-pulse time at the compensation voltage is not known, so the current $\uppi$-pulse time is first measured through Rabi oscillation. The pulse time at the compensation voltage should be shorter than that at an uncompensated voltage, so a voltage scan is performed with a guessed shorter pulse time, for example, one-fourth of the current $\uppi$-pulse time. Even if the pulse time is not equal to the optimal pulse time, the compensation voltage could be found from the symmetry of the Bessel-like curve, otherwise it can be found after adjusting the pulse time a few times depending on the overall excitation level in the scan. Note that the $\uppi$-pulse time is not necessarily the pulse time that yields the best precision of fitting, and a longer pulse time can be more effective due to denser oscillatory pattern of the Bessel-like transition profile. This advantage, nonetheless, will be limited by decaying of the Rabi oscillation, which would reduce the visibility of the oscillatory pattern, and consequently, the fitting precision.

Also, the ions have not been cooled beyond the Doppler cooling, so the maximum transition probability did not reach the unity \bibnote{The Lamb--Dicke parameters for the Raman transition about the principal axes were 0.11, 0.10 (radial), and 0 (axial) in the surface trap and 0.11, 0.05 (radial), and 0 (axial) in the blade trap (the directions of the principal axes are as described in Experimental setup section). For the Doppler cooling, the parameters were 0.058, 0.050 (radial), and 0.099 (axial) in the surface trap and 0.076, 0.033 (radial), and 0.19 (axial) in the blade trap.}. The effect of the thermal motion is reflected in the predicted profile, as in Eq.~(\ref{eq_thermal}), though they still might pose minor discrepancies of measured transition probabilities from fitting. Nevertheless, the discrepancy would not meaningfully lower the performance of estimating the compensation voltage since the effect does not alter the measured Bessel-like profile in the voltage axis and it occurs rather in a symmetric manner about the compensation voltage.

This approach can also allow for monitoring the long-term drift of the null position of the trap. Figure \ref{fig_drift} presents a sequential series of the null scans in time with the carrier transition, which reveals drift of the equilibrium due to laser-induced dielectric charging of the trap chip materials. The amount of drift was approximately 0.03 V in compensation voltage, or 86 V/m in electric field. This deviation from the initial condition is not substantial, but it is still within a detectable range by the method. The standard deviation of the first five scans in this sequential measurement was 0.0016 V, which confirms that the sensitivity in the surface trap estimated from curve fitting is of the same order as the statistical deviation of a repetitive measurement. The repetitive scan was run at a rate of around 3.0 scans per minute, which was sufficiently high for monitoring applications.

\section{Excess micromotion detection by other methods}

The suggested method detects and minimizes the excess micromotion in the direction of the trap's motion-coupled axis which directly affects the motional coherence of the ionic qubits. 
On the other hand, other methods can be used to suppress excess micromotion along their excitation direction by installing additional optical setups if needed, and the measurement results will be presented to facilitate a preliminary comparison.

\begin{figure}[ht]
 \centering
 \includegraphics[width=0.6\textwidth]{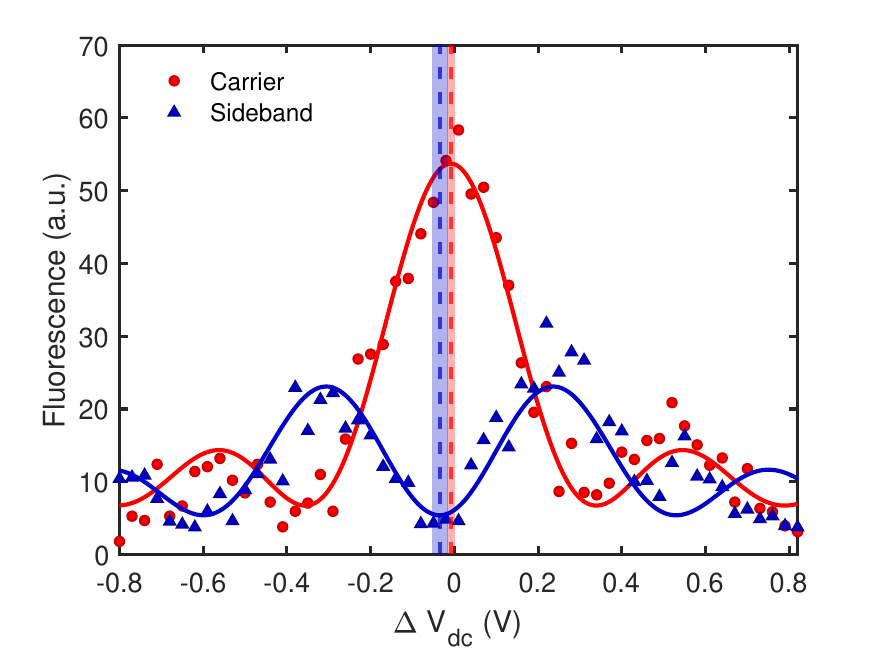}
 \caption{Measurement of ion fluorescence by modified absorption of a weak 935-nm laser. The 95 \% confidence intervals of the estimated compensation voltages are marked by the shaded regions.}
 \label{fig_scan_935}
\end{figure}

First, we can once again utilize direct scanning of the dc voltage but use a sideband-resolved transition by a weak 935-nm repumping laser instead of the Raman transition and measure the intensity of the 369-nm fluorescence in the surface trap. This scheme is used to detect the excess micromotion perpendicular to the chip, and it requires an additional installation of a weak 935-nm beam. The 935-nm laser, injected from the backside of the chip, had an intensity of around 50 mW/cm$^2$ at a trapped ion, which is the minimum intensity to keep the ion trapped. The fluorescence is predicted to follow the discussed Bessel-like behavior, but in the weak-field regime ($\Omega_o \ll \Gamma$) instead of the Rabi oscillation, $|\sin(J_0(\beta)\Omega_o t /2)|^2$ from Eq.~(\ref{bessel}) can be approximated to $|J_0(\beta)\Omega_o t /2|^2 \propto |J_0(\beta)|^2$, where $\Gamma$ is the decay rate of the transition. For a steady state and a fixed detection time, the obtained fluorescence photon count would depend on the magnitude of excess micromotion as $|J_n (\beta)|^2 F_0$ where $F_0$ is the detected photon count when the excess micromotion is minimized, which contains the ion's absorption rate and detection efficiency. The fitting curved was constructed as $a |J_n (\beta)|^2 + d$, where $a$ and $\beta$ are as defined in Eq.~(\ref{eq_thermal}) and $d$ is considered the background signal level. The signal was detected by the PMT with an exposure time of 4 ms. As a demonstration, the result for the carrier transition and the first-order sideband transition ($n=0,1$) of the trap rf frequency is shown in Fig. \ref{fig_scan_935}. The experimental data is well fitted to the predicted curve, with some minor discrepancy which might come from the reality of the setup environment. The estimated sensitivities for the carrier and the sideband transitions could be found in Table \ref{tb_sensitivity_935}. In addition to the overhead of the additional optical setup, the replacement of the original repumping laser with the weak repumping laser also led to inefficient cooling and reduced trapping stability during the measurement.

\begin{table}
\caption{Estimated sensitivity of 935-nm resolved-sideband measurement and photon correlation method to compensation voltage, stray field, and modulation depth.}
\label{tb_sensitivity_935}
\begin{center}
\begin{threeparttable}[b]
\setlength{\arrayrulewidth}{0.22mm}
\renewcommand{\arraystretch}{1.3}
\begin{tabular}{c c c c}
\hline
                     & \multicolumn{2}{c}{935-nm sideband} & \multirow{2}{*}{\begin{tabular}[c]{@{}c@{}}Photon\\ correlation\end{tabular}} \\ \cline{2-3}
                     & Carrier & Sideband \\ \hline
Voltage (V)          & 0.0096  & 0.019    &  (0.0017)\tnote{a} \\ \hline
Electric field (V/m) & 28      & 54       & 1.5 \\ \hline
$\beta$              & 0.067    & 0.13    & - \\ \hline
\end{tabular}
\begin{tablenotes}
    \small
    \item [a] Applied to a farther electrode to tune the E-field in the 369-nm laser direction.
\end{tablenotes}
\end{threeparttable}
\end{center}
\end{table}

\begin{figure}[ht]
 \centering
 \includegraphics[width=0.9\textwidth]{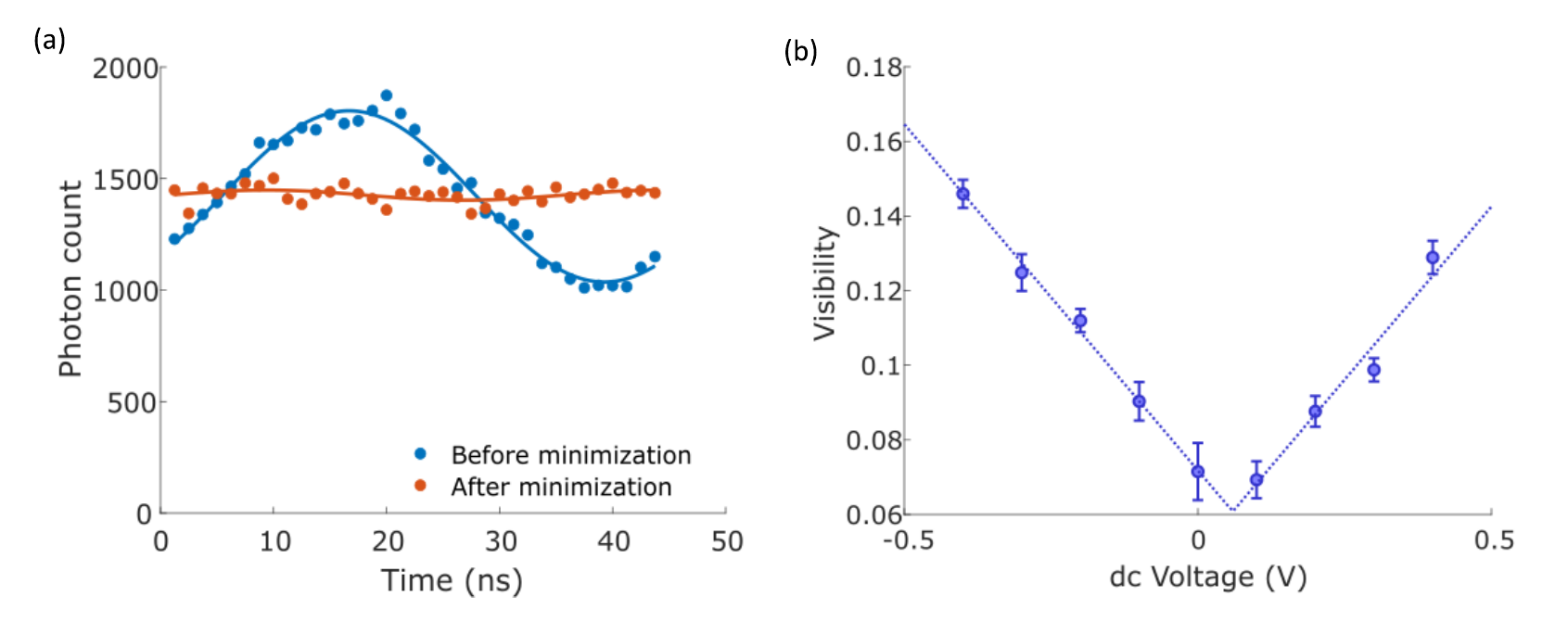}
 \caption{Result of photon correlation measurement. (a) Measured photon count in the period of the rf field before and after the excess micromotion minimization. (b) Visibilities of the photon count in time for various tuning dc voltages.}
 \label{fig_correlation}
\end{figure}

Second, Fig. \ref{fig_correlation} shows the measurement result using the conventional photon correlation method \cite{berkeland_minimization_1998} for excess micromotion along the 369-nm laser direction which is parallel to the surface trap. The technique exploits the modification of absorption in time caused by the Doppler effect from the ion's excess micromotion. In our surface trap setup, excess micromotion in the direction of the 369-nm laser does not cause direct dephasing in quantum control since it is perpendicular to the direction of the controlled motional axis. However, increase of undesired motions may still indirectly degrade the quality of quantum control, and cancellation of the null shift supplements full excess micromotion minimization. When the ion is being Doppler-cooled, the fluorescence is measured with the PMT, and a field programmable gate array (FPGA) estimates the photon arrival time with respect to the phase of the trap rf signal. Since excess micromotion is sinusoidal with the period of the rf field, the measured fluorescence changes sinusoidally as well, and the magnitude of excess micromotion can be estimated from the amplitude of the fitted sinusoidal curve. To create an in-plane electric field with a sufficient amount of component in 369-nm laser beam direction, an outer electrode in the corner side of the surface trap was used to tune the compensation field, unlike other presented measurements. Figure \ref{fig_correlation} shows the correlated photon counts before and after the excess micromotion compensation and the amplitude of the modulation indicates the amount of excess micromotion. The visibility of this modulation was obtained for different dc voltages as shown in Fig. \ref{fig_correlation}, where the uncertainty of the fit could be found in Table \ref{tb_sensitivity_935}. Note that the obtained sensitivity may not represent the best feasible value from the setup and could potentially be enhanced by more precise measurement and additional techniques \cite{lisowski_dark_2005}. In our setup this method can measure the excess micromotion in the plane of the chip, but the measurement perpendicular to the chip requires additional optical setup which significantly interferes with the measurement setup, and generates substantial amount of scattering noise in detection because of the shared wavelength and axis with the detection channel, which makes it practically impossible. Therefore, depending on the specific experimental configuration, both approaches could be employed in a complementary way.

\section{Conclusion}
\label{sec_discussion}

A simple method of detecting and minimizing excess micromotion of a trapped ion using qubit transition and direct scanning of dc voltage has been presented and demonstrated in a surface trap and a blade trap. Measurements are performed in the existing experimental setup for the main qubit control scheme without any need to change the trap environment, which enables diagnosis of excess micromotion in between qubit controls. The method does not require adjustment of the intensity or frequency of the trapping lasers or fields for excess micromotion detection as in certain techniques, which could otherwise potentially compromise the trapping stability. Still, the method provides a decent level of measurement sensitivity, when compared to the reported ones in practical applications that employs conventional techniques. By leveraging the existing setups for excess micromotion detection in other dimensions, this method can ultimately be employed to achieve comprehensive excess micromotion compensation across all dimensions. A necessary requirement for the method is a transition laser having a sufficient amount of momentum, which is unattainable by low-momentum fields such as microwaves. However, many of the commercial quantum computing systems based on trapped ions employ laser-motion coupling to achieve quantum gates by utilizing momentum-coupled transition, so the above requirement is not a barrier for typical systems. On the other hand, our proposed approach can be immediately adopted by existing systems and incorporated into a regular calibration sequence.

\begin{backmatter}
\bmsection{Funding}
Institute for Information \& communications Technology Planning \& Evaluation (IITP) grant (No. 2022-0-01040, IITP-2023-2021-0-01810); The Samsung Research Funding \& Incubation Center of Samsung Electronics (No. SRFC-IT1901-09).

\bmsection{Disclosures}
The authors declare that there are no conflicts of interest related to this article.

\bmsection{Data Availability}
Data underlying the results presented in this paper are not publicly available at this time but may be obtained from the authors upon reasonable request.

\end{backmatter}


\bibliography{micromotion}

\end{document}